\documentclass{jfm}

\usepackage{graphicx}
\usepackage{natbib}
\usepackage{bm}
\usepackage{amsmath}

\ifCUPmtlplainloaded \else
  \checkfont{eurm10}
  \iffontfound
    \IfFileExists{upmath.sty}
      {\typeout{^^JFound AMS Euler Roman fonts on the system,
                   using the 'upmath' package.^^J}%
       \usepackage{upmath}}
      {\typeout{^^JFound AMS Euler Roman fonts on the system, but you
                   dont seem to have the}%
       \typeout{'upmath' package installed. JFM.cls can take advantage
                 of these fonts,^^Jif you use 'upmath' package.^^J}%
      }
  \else
  \fi
\fi

\ifCUPmtlplainloaded \else
  \checkfont{msam10}
  \iffontfound
    \IfFileExists{amssymb.sty}
      {\typeout{^^JFound AMS Symbol fonts on the system, using the
                'amssymb' package.^^J}%
       \usepackage{amssymb}%

      }{}
  \fi
\fi

\ifCUPmtlplainloaded \else
  \IfFileExists{amsbsy.sty}
    {\typeout{^^JFound the 'amsbsy' package on the system, using it.^^J}%
     \usepackage{amsbsy}}
    {}
\fi

\newsavebox{\astrutbox}
\sbox{\astrutbox}{\rule[-5pt]{0pt}{20pt}}

\newcommand{\bfv}{{\bm v}}

\newcommand{\bff}{{\bf f}}
\newcommand{\bfk}{{\bf k}}
\newcommand{\bfx}{{\bf x}}

\newcommand{\bfu}{{\bf u}}
\newcommand{\bfS}{{\bf S}}

\newcommand{\bft}{{\bm \tau}}
\newcommand{\bfn}{{\bf n}}

\title[Acceleration statistics of finite-sized particles in turbulent
flow \ldots ]{Acceleration statistics of finite-sized particles in
  turbulent flow: the role of Fax\'en forces}

\author[E. CALZAVARINI et al.]{E. CALZAVARINI$^1$, R. VOLK$^1$,
  M. BOURGOIN$^2$,\\ E. L\'EV\^ EQUE$^1$, J.-F. PINTON$^1$,
  F. TOSCHI$^3$}

\affiliation{ $^1$Laboratoire de Physique de \'Ecole Normale
  Sup\'erieure de Lyon,\\
  CNRS et Universit\'e de Lyon, 46 All\'ee d'Italie, 69007 Lyon, France.\\
  $^2$Laboratire des \'Ecoulements G\'eophysiques et
  Industriels, \\ CNRS/UJF/INPG UMR5519, BP53, 38041 Grenoble, France.\\
  $^3$Dept. Physics and Dept. Mathematics \& Computer Science, 
   Eindhoven University of Technology, P.O. Box 513, 5600 MB Eindhoven, Netherlands\\
   \vspace{.2cm}
   (\textsc{International Collaboration for Turbulence Research}) 
}  \date{\today} 

\begin{document}
\maketitle

\begin{abstract}
The dynamics of particles in turbulence when the particle-size is larger than the dissipative scale of the carrier flow is studied. 
  Recent experiments
  have highlighted signatures of particles finiteness on their
  statistical properties, namely a decrease of their acceleration
  variance, an increase of correlation times -at increasing the
  particles size- and an independence of the probability density
  function of the acceleration once normalized to their
  variance. These effects are not captured by point particle
  models. By means of a detailed comparison between numerical
  simulations and experimental data, we show that a more
  accurate model is obtained once Fax\'en corrections are
  included.
\end{abstract}

The study of Lagrangian turbulence and of turbulent transport of material particles has received growing interest in recent years (\cite{toschi:2009}).  Modern experimental techniques (based on synchronization of multiple fast cameras, or ultrasonic/laser Doppler velocimetry) allow nowadays to fully resolve particle tracks in turbulent flows (\cite{laporta:2001,mordant:2001,xu:2006,berg:2006,volk:2008a}). These techniques have opened the way toward a systematic study of the dynamics of material (or inertial) particles.  When the particle density is different from the one of the carrier fluid, a rich phenomenology emerges, such as particle clustering and segregation (\cite{squires:1991,calzavarini:2008a,calzavarini:2008b}).   Numerical studies have proven to be essential tools in complementing and benchmarking
early days experimental data: investigations of fluid tracers dynamics have shown remarkable agreement with experiments (\cite{mordant:2004,arneodo:2008,biferale:2008}). Lagrangian numerical studies through Direct Numerical Simulations (DNS) of very small - computationally assumed to be pointwise - particles have also shown encouraging consistency with experimental measurements for inertial particles (\cite{bec:2006,ayyala:2006,salazar:2008,volk:2008b}). However, in many situations the size of the particles is not small with respect to turbulence scales, in particular  the dissipative scale $\eta$.  One example is Plankton which, while neutrally buoyant, cannot be considered as a tracer because of its size in the order of few dissipative scales. Typical marine and atmospherical environmental flows have $\eta \sim O(10) \mu m$. 

The statistics of particle accelerations, which directly reflects the action of hydrodynamical forces, has been used to experimentally assess the statistical signature of  ``large'' spherical particles, i.e. whose diameter $D$ is larger than the
smallest turbulence scale $\eta$. Recent studies~(\cite{voth:2002,qureshi:2007}) and detailed comparison between experiments and numerical simulations~(\cite{volk:2008b}) have shown that  finite-sized neutrally-buoyant particles cannot be modeled as pointwise in numerical studies.   Features which have been clearly associated with a finite particle size are:\\
\noindent i) For neutrally buoyant particles with $D>\eta$ the acceleration variance decreases at increasing the particle size. A scaling law behavior, $\langle a^2 \rangle \sim \varepsilon^{4/3} D^{-2/3}$ (with $\varepsilon$ being the energy
dissipation rate), has been suggested on the basis of Kolmogorov 1941 (K41) turbulence phenomenology (\cite{voth:2002,qureshi:2007}).\\
\noindent  ii) The normalized acceleration probability density function (PDF) depends at best very weakly on the particle diameter. Its shape can be fitted with stretched exponential functions (see \cite{voth:2002,qureshi:2007}).\\
\noindent iii) The autocorrelation function of acceleration shows increased correlation time with increasing particle-size (\cite{volk:2008b}).

While experimentally it is easier to study large ($D > \eta$) particles,  theoretically (and therefore computationally) this turns out a far more difficult task.  Our aim in this article is to study the novel features associated with finite
particles size in developed turbulent flows, while presenting an improved numerical model capable to solve most of the discrepancies between experiments and simulations noticed in (\cite{volk:2008b}). We show that qualitatively and quantitatively the new features are well captured by an equation of motion which takes into account the effect of the nonuniformity of the flow at the particle-scale. To our knowledge the impact on acceleration statistics of such forces, known since a long time as Fax\'en corrections (\cite{faxen:1922}) has never been considered.

The article is organized as follows. First we comment on the problems of obtaining an equation of motion for finite-sized particles. We examine the approximation on which point-particles equations rely and discuss two highly simplified models for the dynamics of small ($D<\eta$) and finite-sized ($D>\eta$) particles. Section \ref{sec2} gives the numerical implementation of the proposed Fax\'en-corrected model.  In section \ref{sec3} we show basic physical differences between the statistics of particle acceleration given by numerics with or without Fax\'en corrections.  Section \ref{sec4} contains the comparison of the model against experimental results, focusing on neutrally buoyant particles. Finally in Sec. \ref{sec5} we summarize the results, we critically review the model and discuss how it can be improved.

\section{Equation of motion for finite-sized particle in turbulence}
Many studies on fine particulate flows have based particle's description on an equation - referred to as Maxey-Riley-Gatignol - which is an exact derivation of the forces on a particle in a nonuniform unsteady flow in the limit of vanishing Reynolds numbers $Re_p=D v_s/\nu$ and $Re_{S} = D^2 \Gamma / \nu$, where $v_s$ is the slip particle velocity respect to the fluid and $\Gamma= |\nabla \bfu|$ the typical shear-scale in the flow (\cite{maxey:1983,gatignol:1983}).  In the net hydrodynamical force acting on a particle given by this equation one recognizes several contributions: the steady Stokes drag, the fluid acceleration force (sum of the pressure gradient and the dissipative forces on the fluid), the added mass, the buoyancy, the history Basset-Boussinesq force, and Fax\'en corrections.  When the control parameters $Re_{p}$ and $Re_{S}$ become finite, the non-linearity of the flow dynamics in the vicinity of the particle must be taken into account (see the review by \cite{michaelides:1997}).  An expression for the added mass term which is correct at any $Re_p$ value has been derived by \cite{auton:1988}.  But much more complicated is the situation for the other forces involved.  The drag term becomes $Re_{p}$ dependent and empirical expressions based on numerical computations have been proposed (see \cite{clift:1978}). Furthermore,  a lift force appears at finite values of $Re_{p}$ and
$Re_{S}$. This force is notably hard to model because of the non-linear combination of shear and vorticity, and approximate expressions based on Saffman (small $Re_p$) and Lighthill-Auton (large $Re_p$) mechanisms are often used in studies (see e.g. discussion on lift on bubbles by \cite{magnaudet:1998}).

Theoretical and numerical studies of fine disperse multi-phase flows, which aim at describing the behavior of a large number of particles, have adopted simplified models where the sub-dominant terms in Maxey-Riley-Gatignol equation are neglected (\cite{balkovsky:2001,bec:2005}).  A minimal model, used to address particle Lagrangian dynamics in highly turbulent suspensions, takes into account only a few ingredients: the Stokes drag, the Auton added
mass and the fluid acceleration term (\cite{babiano:2000,calzavarini:2008a}).  This leads to:
\begin{equation}\label{ppmodel}
  \frac{d \bfv}{dt} =  \frac{3\ \rho_f}{\rho_f + 2\ \rho_p} \ \left(  \frac{D\bfu}{Dt}  + \frac{3 \nu}{a^2} \left(  \bfu - \bfv \right) \right),
\end{equation}
where $\rho_f$ and $\rho_p$ are respectively the fluid and the particle density, $\nu$ the fluid kinematic viscosity and $a$ the radius of the particle, which is considered spherical.  However, the size of the particle in the above equation is essentially {\it virtual}.  Equation (\ref{ppmodel}) contains only a time-scale, namely the particle relaxation time $\tau_p$, which embodies a particle length-scale merely in combination with the kinematic viscosity of the flow and with the densities coefficients, i.e.,  $\tau_p \equiv a^2 (\rho_f + 2\ \rho_p) / (9 \nu \rho_f)$. In practice, the drag term in equation (\ref{ppmodel}) performs a purely temporal filtering of the flow velocity fluctuations.

It is the role of Fax\'en terms to account for the non-uniformity of the flow at the particle-size.  Fax\'en forces represent necessary physical corrections when analyzing the behavior of $D>\eta$ particles in turbulence.   The Fax\'en theorem for the drag force on a moving sphere states the relation 
\begin{equation}\label{fd}
\bff_{D} =  6 \pi \nu \rho_f a\  \left(  \frac{1}{4 \pi a^2} \int_{S_a} \bfu(\bfx)\ d\bfS  -  \bfv \right) =
                      6 \pi \nu \rho_f  a \left(  \langle \bfu \rangle_{S_a}  -  \bfv \right),
\end{equation}
where the integral is over surface of the sphere and $\bfu(\bfx)$ the nonhomogeneous steady motion  of the fluid in the absence of the sphere. As later shown by \cite{gatignol:1983}, Fax\'en force corrections via sphere volume averages should also be included on the inertial hydrodynamic forces acting on the sphere. In particular the expression for the fluid acceleration and added mass force becomes:
\begin{equation}\label{fa}
\bff_{A} =  \frac{4}{3}\pi a^3 \rho_f \left(   \langle \frac{D \bfu}{Dt} \rangle_{V_a} + \frac{1}{2} \left( \langle  \frac{d \bfu}{dt} \rangle_{V_a} - \frac{d \bfv}{dt}    \right) \right)
\end{equation}
where similarly as above $\langle \ldots \rangle_{V_a}$ denotes the volume average over the spherical particle.
Putting together the two force  contributions of Eqs. (\ref{fd})-(\ref{fa}) into an equation of motion for a sphere, 
$(4/3)\pi a^3 \rho_p  \ d \bfv/dt = \bff_{D} + \bff_{A}$,
and keeping into account the Auton added mass correction for finite $Re_P$, i.e., $d \bfu/dt  \to D \bfu/Dt$, we obtain the phenomenological Fax\'en-corrected equation of motion:
\begin{equation}\label{fcmodel}
  \frac{d \bfv}{dt} =  \frac{3\ \rho_f}{\rho_f + 2\ \rho_p} \ \left(  \langle \frac{D\bfu}{dt} \rangle_{V_a} + \frac{3 \nu}{a^2} \left(  \langle \bfu \rangle_{S_a} - \bfv \right) \right).
\end{equation}

In the small particle limit, when $\bfu \simeq \bfv$, corrections
can be approximated by Taylor expansion $\langle \bfu(\bfx,t)
\rangle_{S_a} \simeq \bfu + \frac{a^2}{6}\nabla^2 \bfu + O(a^4)$;
$\langle \frac{D \bfu(\bfx,t)}{Dt} \rangle_{V_a} \simeq \frac{d}{dt}
\left( \bfu + \frac{a^2}{10}\nabla^2 \bfu + O(a^4) \right)$, therefore
the first order Fax\'en correction accounts for the curvature of the
unperturbed flow at the particle location.  In a turbulent flow the
correction term becomes important when $a > \eta$, with a weak
Taylor-based Reynolds number $Re_{\lambda}$ dependence.  With the assumption
$\nabla^2 \bfu \sim 15 \bfu_{rms}/\lambda^2 $, where $\lambda$ is the
Taylor microscale, one can roughly estimate $a^2 \nabla^2 \bfu \sim
15 \bfu_{rms}\ (a/\lambda)^2 \sim \bfu_{rms}\ (a/\eta)^2
\sqrt{15}/Re_\lambda$.

\section{Numerical implementation of particle model and tubulence DNS}\label{sec2}
We adopt here a further approximation which allows efficient numerical computations of Eq. (\ref{fcmodel}).  Volume
averages at particles' positions are substituted by local interpolations after filtering by a Gaussian envelope with standard
deviation, $\sigma$, proportional to the particle radius. Gaussian
convolutions are then efficiently computed in spectral space, and the
Gaussian volume averaged field reads:
\begin{equation}
  \langle u_i \rangle_{G,V_\sigma}(\bfx) =  \mathcal{ DFT}_{(N^3)}^{-1} \left[  \tilde{G}_{\sigma}(\bfk)\  \tilde{u_i}(\bfk) \right],
\end{equation}
where $\mathcal{ DFT}_{(N^3)}^{-1}$ denotes a discrete inverse Fourier transform on a grid $N^3$, $\tilde{G}_{\sigma}(\bfk) = \exp({-\sigma^2 \bfk^2}/2) $ is the Fourier transform of a unit volume Gaussian
function of variance $\sigma$ and $\tilde{u_i}(\bfk)$ is the Fourier
transform of a vector field (the material derivative of fluid velocity
in Eq. (\ref{fcmodel})).  The surface average is obtained using the exact relation:
\begin{equation}\label{fmodel}
  \langle u \rangle_{S_a} = \frac{1}{3 a^2} \frac{d}{da} \left(  a^3  \langle \bfu \rangle_{V_a} \right),
\end{equation}
which leads to:
\begin{equation}
\langle u_i \rangle_{G,S_\sigma}(\bfx) =  \mathcal{ DFT}_{(N^3)}^{-1} \left[  \tilde{S}_{\sigma}(\bfk)\  \tilde{u_i}(\bfk)  \right],  
\end{equation}
where: $ \tilde{S}_{\sigma}(\bfk) = \left( 1 - \frac{1}{3} \sigma^2
  \bfk^2 \right) e^{-\frac{1}{2} \sigma^2 \bfk^2}$.  It can be shown
that with the choice $\sigma=a/\sqrt{5}$, the Gaussian convolution
gives the right prefactors for the Fax\'en correction in the limit $a
\to 0$.  Our simplified approach for the integration of
(\ref{fcmodel}) (FC model) allows to track inertial particles in
turbulent flows with minimal additional computational costs as
compared to Eq. (\ref{ppmodel}) (PP model): the fluid acceleration and
velocity fields are filtered once for every particle radius size, then
the averaged flow at the particle positions are obtained through a
tri-linear interpolation.  We track particles via Eq. (\ref{fcmodel}) in
a stationary homogeneous isotropic flow, generated by large-scale
volume forcing on a cubic domain. The Navier-Stokes equation is
discretized on a regular grid, integrated using a pseudo-spectral
algorithm, and advanced in time with a $2^{\rm nd}$ order Adams-Bashford
integrator.

We have explored in a systematic way the two-dimensional parameter
space $[\rho_p/\rho_f, D/\eta]$ in the range $\rho_p/\rho_f \in
\left[0.1, 10 \right]$ and $D/\eta \in \left[2, 50 \right]$  for a
turbulent flow at $Re_{\lambda}=180$ ($512^3$ collocation points). We
tracked $\sim 2\cdot 10^6$ particles for a total of $\sim 4$
large-eddy turnover time in statistically stationary conditions. Lower
resolution DNS at $Re_{\lambda}=75$ $(128^3)$ have been used to
explore a larger parameter space and to study the differences between
the point-particle model (\ref{ppmodel}) and the Fax\'en corrected
model (\ref{fcmodel}) in the asymptote $D\to L$ (with $L$ the
turbulence integral-scale). The validation of the numerics have been
performed through careful comparison with an independent code
implementing the same algorithm for particles, but with different
forcing scheme, temporal integration method (Verlet algorithm), and
local interpolation scheme (tri-cubic algorithm).

%%%%%%%%%%%%%%
\begin{figure}
  \begin{center}
    \includegraphics[width=0.95\columnwidth]{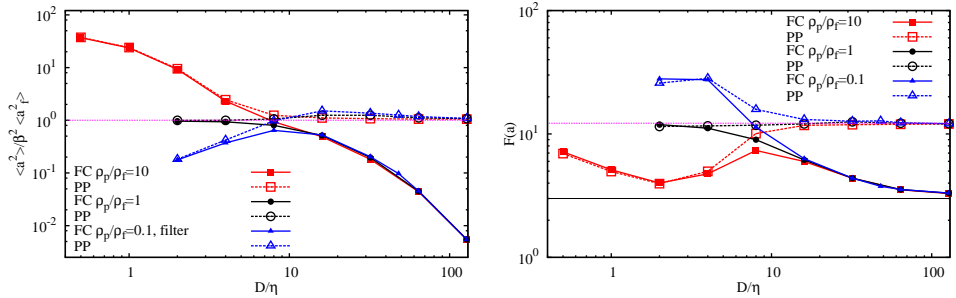}
    \caption{(left) The acceleration variance $\langle a^2 \rangle$ normalized by $\beta^2 \langle a^2_f \rangle$   
    versus the particle diameter as derived from the Fax\'en-Corrected model (solid
      lines/symbols), and from Point-Particle the model system (dashed
      lines/empty symbols). Density ratios shown are
      $\rho_p/\rho_f = 0.1, 1, 10$, i.e., heavy ($\square$), neutral ($\circ$) and light ($\triangle$)
      particles. (right) Same as above for the acceleration flatness
      $F(a) = \langle a^4 \rangle / \langle a^2 \rangle^2$. Horizontal
      lines shows the flatness of the fluid acceleration $F(a_f)$ and
      the flatness value for Gaussian distribution $F(a)=3$. Data
      from simulations at $Re_{\lambda}=75$.}
    \label{figure1}
  \end{center}
\end{figure}
%%%%%%%%%%%%%%%

\section{Phenomenology of Point-Particle and Fax\'en-Corrected models}\label{sec3}
We compare the statistics of acceleration of particles tracked via the PP and FC equations.  In the small particle limit ($D/\eta \to 0$) the two model equations behave the same way and the particle trajectory becomes the one of a fluid
tracer. The ensemble-average acceleration variance reaches the value $\langle a^2 \rangle \to \langle a^2_f \rangle$ with the subscript $f$ labeling the fluid tracer acceleration. As the particle diameter is increased we
notice important differences between the two models. In the PP model the drag term becomes
negligible and one gets $\langle a^2 \rangle \simeq \beta^2 \langle a^2_f \rangle$, with $\beta = 3 \rho_f/(\rho_f + 2 \rho_p)$. In the FC model the volume average of the fluid acceleration $D\bfu/Dt$ reduces progressively the particle
acceleration. This is illustrated in Figure \ref{figure1} (left), where the particle acceleration variance (normalized by $\beta^2 \langle a_f^2 \rangle$)  is shown for three cases: neutral buoyant, heavy ($\rho_p/\rho_f=10$) and light
($\rho_p/\rho_f=0.1$) particles.  We note that the behavior of $\langle a^2 \rangle$ for particles whose diameter is roughly larger than $10 \eta$ seems to be identical apart from the scaling factor $\beta^2$.

Differences are also present in higher order moments: for this we
focus on the flatness $F(a) \equiv \langle a^4 \rangle/ \langle a^2
\rangle^2$. In the large $D$ limit PP model gives the rather
unphysical behavior $F(a) \simeq F(a_f)$, that is to say large
particles, irrespectively of their density, show the same level of
intermittency as a fluid tracer. On the other hand the Fax\'en
corrected equation gives asymptotically $F(a) \simeq 3$, i.e., the
Gaussian flatness value, meaning that acceleration of large particles
independently of their mass density value has lost its intermittent
character, see Fig. \ref{figure1} (right).  Furthermore, it is
noticeable that above a certain critical value of the diameter the
flatness of heavy/neutral and light particles reaches the same level:
this suggests that also the PDFs may have very similar shapes.

\section{Comparison with experiments}\label{sec4} 
We study now how the FC model compares with  the experimental observations listed in the introduction -- recalling than none  is captured by the PP model.

%%%%%%%%%%%%%%%%%%
\begin{figure}
  \begin{center}
    \includegraphics[width=0.7\columnwidth]{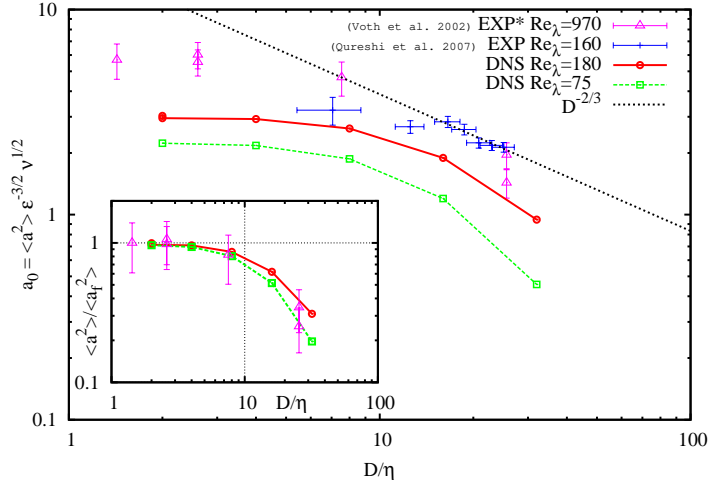}
    \caption{One component acceleration variance versus particle
      size. Acceleration is normalized by the Heisenberg-Yaglom
      relation, while the particle size is is normalized by the 
      dissipative scale. DNS results have uncertainty of the order of
      the symbol size. Data EXP are from \cite{qureshi:2007}, with EXP* measurement from \cite{voth:2002} (Figure 32) --  particles with
      density contrast $\rho_p/\rho_f=1.06$.  Inset: $\langle
      a^2 \rangle/\langle a_f^2 \rangle$ vs. $D/\eta$ from the same
      DNS and EXP* measurements.}
    \label{figure2}
  \end{center}
\end{figure}
%%%%%%%%%%%%%%%%%%%

\subsection{Acceleration variance}
%i) 
In figure \ref{figure2} the behavior of the one-component acceleration
variance, normalized by the Heisenberg-Yaglom scaling, $a_0
= \langle a_i^2 \rangle \epsilon^{-3/2} \nu^{1/2}$, is displayed.
Although this way of normalizing the acceleration has a weak Reynolds
number dependence (see \cite{voth:2002,bec:2006}) we notice a very
similar behavior as compared to the experimental measurements at
$Re_{\lambda}=160$ by \cite{qureshi:2007} and with $Re_{\lambda}=970$
experiments by \cite{voth:2002}.  In the inset of Figure \ref{figure2}
the same quantity but with a different normalization is shown. The
particle acceleration variance there is divided by the second moment of
fluid tracer acceleration $\langle a^2_f \rangle$. The experimental
data from \cite{voth:2002} can also be rescaled in the same way by
dividing $a_0$ by the value for the smallest considered particle
(which has size $D \simeq 1.44 \eta$ and essentially behaves as a
fluid tracer). This alternate way of looking at the data renormalizes the
weak $Re_{\lambda}$ dependence, providing a good agreement between the
DNS and experiments even when comparing results with one order of
magnitude difference in $Re_{\lambda}$.

In a DNS one can estimate the relative weight of the terms
contributing to the total acceleration: the drag and fluid
acceleration terms, respectively: $a^D=( \langle \bfu \rangle_S - \bfv
)/\tau_p$ and $a^A=\beta \langle \frac{D \bfu}{Dt} \rangle_V$.  It is
important to note that in the case of neutrally buoyant particles, one
finds $\langle a \rangle_{rms} \simeq \langle a^A \rangle_{rms}$ with
percent accuracy. It indicates that the observed effect - decrease 
of particle acceleration variance for increasing particle diameter -
comes uniquely from volume averaging of fluid acceleration at the particle
position.  The drag contribution is sub-leading at all $D$ values
(from few percent up to 15\% of total acceleration variance); it
just contributes to compensate the $a^D a^A$ correlations.  Stated
differently, one can say that the acceleration of a finite-size
neutrally buoyant particle is essentially given by $\langle D\bfu/Dt
\rangle_{V_a} = \langle \nabla \cdot \bft + \bff_e \rangle_{V_a}
\simeq \frac{1}{3a} \langle \bft \cdot \bfn \rangle_{S_a}$, where
$\bft$ is the stress tensor, $\bfn$ a unit norm vector pointing
outward the sphere and $\bff_e$ the external large-scale forcing whose
contribution $\langle \bff_e \rangle_{V_a} \simeq 0$ is negligible at
the particle scale. One expects the situation to be different for
particles whose density does not match that of the fluid.

Our simulations are consistent with the $ a_0 \sim D^{-2/3}$ scaling
which has been proposed on the basis of dimensional arguments rooted
on Kolmogorov 1941 turbulence phenomenology without special
assumptions on particle dynamics (\cite{voth:2002,qureshi:2007}); 
however at  $Re_{\lambda}=180$ the  scale-separation is still too limited to observe a true scaling range.

%%%%%%%%%%%%%%%%%%%
\begin{figure}
  \begin{center}
    \includegraphics[width=0.7\columnwidth]{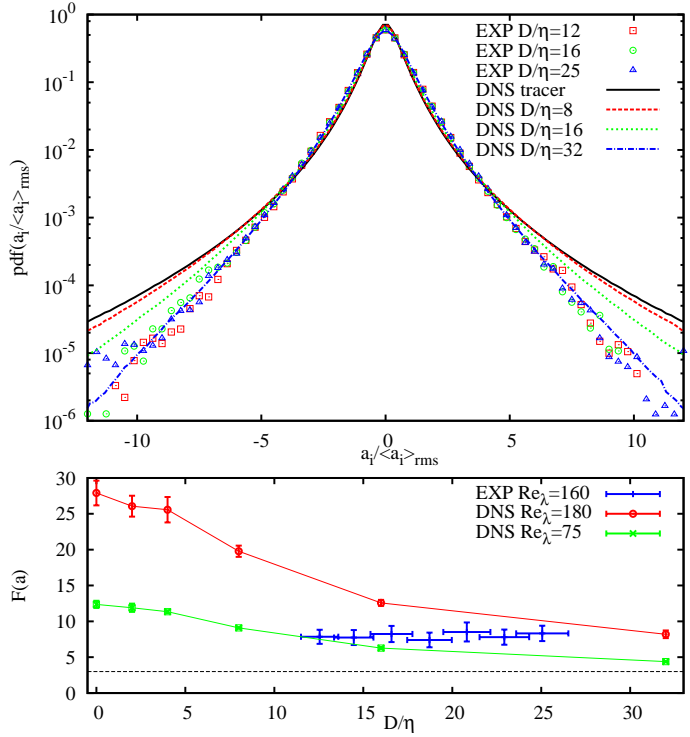}
    \caption{(top) Comparison of probability density functions of
      acceleration normalized by its \textit{rms} value, from EXP
      \cite{qureshi:2007} at $Re_{\lambda}=160$ and DNS at
      $Re_{\lambda}=180$. (bottom) One component acceleration flatness
      $F(a) = \langle a_i^4 \rangle/\langle a_i^2 \rangle^2$ versus
      the normalized particle diameter, $D/\eta$, from the same experiment and DNS at two
      different Reynolds numbers.}
    \label{figure3}
  \end{center}
\end{figure}
%%%%%%%%%%%%%%%%%%%

\subsection{Acceleration probability density function}
The second quantity under study is the acceleration probability density
function. Here, to cope with $Re_{\lambda}$ effects, one compares only
the two most similar data sets: the DNS at $Re_{\lambda}=180$ and the
experiment at $Re_{\lambda}=160$ \cite{qureshi:2007}.  Experiments
have revealed a universal behavior for acceleration PDF normalized by
$\langle a_i^2 \rangle^{1/2}$ in the size-range $D=12 - 25 \eta$.  DNS
instead shows a systematic difference in the their trend: larger
particles have less intermittent acceleration statistics, see Figure
\ref{figure3} (top). However, the shape of the PDF in the limit of
large particles $D \simeq 30 \eta$ shows a good similarity.  To
better visualize differences, in Figure \ref{figure3} (bottom),
we show the flatness $F(a)$ vs. particle diameter for DNS and
experiments. As already observed, the FC model leads to decreasing
intermittency for bigger neutral particles, and in the asymptotic
limit $(D \to L)$ to Gaussian distribution; also acceleration flatness
is an increasing function of $Re_{\lambda}$.  Qureshi and coworkers'
experiment on the other hand shows a $D$-independent behavior around
$F(a) =8.5$.
A further possible source of differences can be connected to the variations in the large scale properties of turbulent flows: Experimental tracks come from a decaying grid-generated turbulence, simulations instead uses volume large-scale forced flow in a cubic domain without mean flow. 

\subsection{Acceleration time-correlation}
\begin{figure}
  \begin{center}
    \includegraphics[width=0.7\columnwidth]{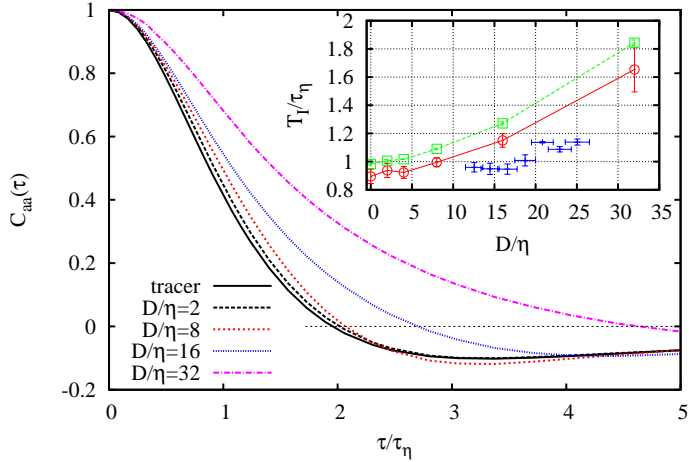}
    \caption{
      Autocorrelation function of acceleration $C_{aa}(\tau)$
      for neutral particles $(\rho_p=\rho_f)$, with different sizes $D= 2,8,16,32 \eta$ and a tracer particle. Inset:  integral acceleration time: $T_I = \int_0^{T_0} C_{aa}(\tau) \ d\tau$ with $T_0$ the zero-crossing time $C_{aa}(T_0)=0$,  versus particle diameter. Symbols: ($\square$, $\circ$) for DNS at  $Re_{\lambda}= (75, 180)$, data ($+$) from experiments at $Re_{\lambda}=160$.}
     \label{figure4}
  \end{center}
\end{figure}

Finally, we consider the dynamics of the neutral particles. We study the normalized one-component correlation function, 
$C_{aa}(\tau) \equiv \langle a_i(t)a_i(t+\tau)\rangle/\langle a_i^2\rangle$.
In \cite{volk:2008b} it has been noted that PP
model can not account for the increasing autocorrelation for larger
particles. This is understood from equation
(\ref{ppmodel}): In the large $D$ limit the drag term is negligible
and the acceleration of a neutrally buoyant particle is dominated by
the inertial term $D \bfu / Dt$. Therefore the time-correlation of
acceleration, $C_{aa}(\tau)$, is related to the temporal correlation of
$D \bfu / Dt$ along the particle trajectory. Because in the
large $D$ limit $\bfv \neq \bfu$ (\cite{babiano:2000}), 
one expects an acceleration correlation time which is equal or even shorter than
the one of a fluid tracer. This is confirmed by our numerics
based on the PP equation (\ref{ppmodel}).  In the FC model
instead, the averaged quantity $\langle D \bfu / Dt \rangle_{V_a}$
dominates the particle's acceleration and so its time correlation
$C_{aa}(\tau)$. In Figure \ref{figure4} we show that simulations based on Eq. (\ref{fcmodel}) display increasing correlation time for bigger particles, as observed in experiment
(\cite{volk:2008b}, although at much larger $Re_{\lambda}$ values). 
A detailed comparison of $C_{aa}(\tau)$ curves coming from DNSs with experiments by Qureshi and coworkers is at present not possible, because of limited statistics. 
Therefore, we examine integral quantities such as an integral acceleration time, $T_I$.  Since by kinematic constraint the time integral of $C_{aa}(\tau)$ for a small tracer is zero, we  define $T_I$ as the integral over time of the positive part of $C_{aa}(\tau)$; this choice proves to be stable in the experiments and weakly dependent on the unavoidable (gaussian) smoothing of noisy data sets (see \cite{volk:2008a}).  
The result of this analysis is reported in figure \ref{figure4} (inset). The order of magnitude of $T_I/\tau_{\eta}$, which is very near unity,  as well its increasing trend with $D$ qualitatively  confirms the prediction of the FC model at similar Reynolds number. Using DNS results, it is also interesting to note that this time decreases with increasing Reynolds number.

%%%%%%%%%%%%%%%%%%%%%
%%%%%%%%%%%%%%%%%%%%%

\section{Discussion of results and conclusions}\label{sec5} 
We have investigated the origin of several experimental observations
concerning neutrally-buoyant finite-size particle acceleration in
turbulent flows and shown the relevance of Fax\'en corrections.
Fax\'en terms account for inhomogeneities in the fluid flow at the
spatial extension of the particle. They act as spatial coarse-graining
of the surrounding turbulent flow, in contrast with the drag
term which performs a temporal filtering. Numerically, the spatial
average is efficiently implemented via Gaussian filtering in spectral
space.  Comparing with experimental measurements, the main
achievements of the Fax\'en-corrected model are: (i)
prediction of the reduction of acceleration fluctuations at increasing
the particle size; (ii) prediction of the increasing of acceleration
time correlation at increasing the particle size. Both effects
originate from the volume average of the fluid acceleration term, or
in other word from the surface average of the stress tensor of the
unperturbed flow.  While giving the correct trend, the FC model does
not solve the puzzling point of invariant PDF with particle size,
observed by \cite{qureshi:2007}.

The Fax\'en corrected model marks a substantial improvement in the statistical description of realistic turbulent particle suspensions.  We emphasize that none of the observed trends in the acceleration of neutrally buoyant particles can be captured by
purely local models, as e.g. the point-particle one in eq. (\ref{ppmodel}).
Fax\'en corrections are of highest relevance in the case of
neutrally buoyant particles, because it is the case for which the slip
velocity ($v_s \equiv |\bfv - \langle \bfu \rangle_S |$) is the smallest
(as compared to $\rho_p \neq \rho_f$ particles) and therefore where
drag, history , lift have the least impact on the net force.  In our
case we observe that when increasing the size of particles, the PDFs
of slip velocity normalized by the fluid velocity rms value
($v_s/\bfu_{rms}$) change from a sharp delta-like shape (for tracers) to
larger distributions approaching a Gaussian (for large
particles).  A size-dependent slip velocity for neutrally buoyant
tracers in chaotic flows has been reported recently in
\cite{oullette:2008}: Fax\'en corrections to the added mass should be
significant in that case too.
We also observe that the particle-Reynolds number $Re_p$ measured in our 
simulations attains values in the order of the hundreds, hence a more accurate description of the drag coefficient (see for instance \cite{clift:1978}) may be important particularly for a faithful reproduction of the  far tails of the acceleration PDF.

\textit{Acknowledgments:} The authors acknowledge the Grenoble team (N. M. Qureshi,
C. Baudet, A. Cartellier and Y. Gagne) for generously sharing their
experimental data measurements, and J.Bec for useful discussions. 
Numerical simulations were performed at SARA (The Netherlands), CINECA (Italy) and PSMN, ENS-Lyon (France).  
Numerical raw data on FC particles are freely available on iCFDdatabase
(http://cfd.cineca.it) kindly hosted by CINECA (Italy).

%\bibliographystyle{jfm}
%\bibliography{biblio_fsm}

\end{document}